\documentclass[10pt,letterpaper]{article}
\usepackage[utf8]{inputenc}

\usepackage{amsfonts,amsmath,latexsym,amssymb}
\usepackage{authblk}
\usepackage{statmath}
\usepackage[colorlinks=true,
allcolors=blue]{hyperref}
\usepackage{multirow}
\usepackage{multicol}
\usepackage{url}
\usepackage{color}
\usepackage{geometry}
\usepackage{graphicx}
\usepackage{subfigure}
\usepackage{stmaryrd}
\usepackage{caption}
\usepackage{indentfirst} 
\usepackage{tabularx, booktabs}
\usepackage{float}
\usepackage[table]{xcolor}
\usepackage{notes2bib} 
\usepackage{setspace}
\usepackage[numbers,sort,compress]{natbib}
\usepackage{etoolbox} 
\patchcmd{\thebibliography}{\section*{\refname}}{}{}{} 

\begin{document}
	
	\title{Physics-based tissue simulator to model multicellular systems: A study of liver regeneration and hepatocellular carcinoma recurrence}
	
	\author[1]{Luciana Melina Luque}
	\affil[1]{Instituto de Física de Líquidos y Sistemas Biológicos - CONICET. La Plata, Argentina.}
	
	\author[1,2]{Carlos Manuel Carlevaro}
	\affil[2]{Departamento de Ingenier\'ia  Mec\'anica, Universidad Tecnol\'ogica Nacional, Facultad Regional La Plata, La Plata, Argentina.}
	
	\author[3]{Camilo Julio Llamoza Torres}
	\affil[3]{Hospital Clínico Universitario Virgen de la Arrixaca. Murcia, España}
	
	\author[4]{Enrique Lomba}
	\affil[4]{Instituto de Química Física Rocasolano - CSIC. Madrid, España}
	
	\date{}

	\maketitle
	\begin{abstract}
		
		We present a multiagent-based model that captures the interactions between different types of cells with their microenvironment, and enables the analysis of the emergent global behavior during tissue regeneration and tumor development. Using this model, we are able to reproduce the temporal dynamics of regular healthy  cells and cancer cells, as well as the evolution of their three-dimensional spatial distributions. By tuning the system with the characteristics of the individual patients, our model reproduces a variety of spatial patterns of tissue regeneration and tumor growth, resembling those found in clinical imaging or biopsies. In order to calibrate and validate our model we study the process of liver regeneration after surgical hepatectomy in different degrees. In the clinical context,  our model is able to predict the recurrence of a hepatocellular carcinoma after a $70\%$ partial hepatectomy. The outcomes of our simulations are in agreement with experimental and clinical observations. By fitting the model parameters to specific patient factors, it might well become a useful platform for hypotheses testing in treatments protocols.
		
	\end{abstract}

	\section{Author summary}
	\label{sec:author}
	
	We introduce an off-lattice agent-based model to simulate tissue-scale features that emerge from basic biological and biophysical cell processes. In order to calibrate and validate our model, we have considered the liver regeneration response after a $30\%$ partial hepatectomy in which the liver recovers its original volume due to the hypertrophy of the hepatocytes. Subsequently, we have modeled the same process but after a $70\%$ partial hepatectomy, in which the liver recovers its original volume due to the hypertrophy and the proliferation of the hepatocytes. Unfortunately, the precise mechanisms of initiating, promoting and terminating regenerative responses remain unknown. As a consequence, we have proposed a modeling approach in which such processes are regulated by a hypothetical substrate that diffuses in the cell microenvironment. As a further test, we have, in one hand, implemented our model to predict the liver response after a $50\%$ partial hepatectomy and, on the other hand, explored our model's ability to account for the recurrence of a hepatocellular carcinoma. The outcomes of our simulations agree with experimental data and clinical observations, which comes to underline the significant descriptive and predictive power of this computational approach. Even though our model needs to be further extended to incorporate patient specific clinical data, these results are a promising step in the direction of  a personalized estimation of tissue dynamics from a limited number of measurements carried out at diagnosis.
	
	\section{Introduction}
	\label{sec:introduction}
	
	Many significant multicellular system problems such as tissue engineering, evolution in bacterial colonies, and tumor metastasis can only be understood by studying how individual cells grow, divide, die, move, and interact. Tissue-scale dynamics emerges as cells are influenced by biochemical and biophysical signals in their microenvironment, bearing in mind that cells themselves continually remodel their own microenvironment. Thus, the ideal scenario to study a multicellular system's biology must simultaneously address: tissue microenvironments with multiple diffusing chemical signals (e.g., oxygen, drugs, and signaling factors), and the dynamics of many mechanically and biochemically interacting cells.
	
	To that aim, mechanistic dynamical systems models described by ordinary differential equations have been developed \cite{cc16, cc17, cc18, cc19, cc20, cc21, cc22, cc23, cc24, cc25} (cf. those reviewed in \cite{cc26} and \cite{cc27}). Even though these models are very useful, they lack the spatial resolution that would enable the examination of intratumoral heterogeneity and its correlation with treatment efficacy. This is a relevant feature since intratumoral heterogeneity has become a central element for understanding important aspects of cancer treatment such as drug resistance and biomarkers \cite{cc28, cc29}. 
	
	A widely used modeling paradigm in the study of complex biological systems is the \textit{agent-based model} (ABM) \cite{ABM2, ABM1}. ABMs are implemented mainly to simulate the actions, behaviors and interactions of autonomous individual or collective entities, with the aim of exploring the impact of an agent or a type of behavior in the system. Agents are independent units trying to accomplish a set of goals. The purpose of an ABM is to explore variations in the system behavior due to agent characteristics or rules. These attempt to emulate the general behavior of the system and predict the patterns of complex phenomena. Agents behave independently, but they react to the environment, modify system properties, and incorporate new  agents. They also have the ability to ``learn'', that is, to avoid previously unsuccessful decisions and favor successful ones, as well as to ``adapt'', i.e. change their behavior in response to variations  of the properties of the system. Their basic advantage is that they provide predictive power on a large scale \cite{cc44}.
	
	In many cases the system being modelled is usually comprised of millions of components. In those cases a great level of abstraction/simplification must be applied to the model to render it useful \cite{abmsim4} (we refer the reader to \cite{abmsim3} for as extensive discussion on the accuracy of computational models). For that reason, once a preliminary model has been constructed it must be subjected to verification, calibration and validation. Verification is the process of determining how accurately prior knowledge and underlying assumptions have been translated into mathematical form. Calibration is the process by which parameters in a model are adjusted so as to match model performance to experimental data. Finally, model validation is the process of evaluating model performance against the primary design goal. In the case of biological models, this usually aims at achieving a close match between model and experiment \cite{abmsim2}.
	
	In a medical context, ABMs should allow simulating clinical trials in sufficient detail to study the subject's response to changes in therapy in simulations, rather than in patients. ABM have been used to study many different aspects of cancer, including tumor growth, cancer cell migration, metabolism, evolutionary dynamics, metastasis, angiogenesis and the role of cancer stem cells \cite{cc30, cc31, cc32, cc33, cc34, cc35, cc36, cc37,cc38, cc39, cc42, cc43, physicell, chichang}.  The reader is referred the extensive review by Norton and coworkers \cite{nortonreview} for a more detailed presentation of the potential of ABMs in the context of cancer modeling.
	
	Here, we will present a mechanistic off-lattice agent-based model to simulate tissue-scale features that emerge from basic biological and biophysical cell processes. This ABM will be validated and put to test for  modeling the unusual ability of the liver to regenerate \cite{taub}. 
	Even when $70\%$ of its mass is surgically removed, the remnant portion expands to compensate for the lost tissue and functions \cite{miya7}. The multilobular structure of the liver allows the surgical resection of a lobe of choice to achieve different degrees by partial hepatectomy (PH). Because the resection of lobes does not induce damage to the remaining tissue, PH is a clean model. Therefore, liver regeneration after PH has long been an excellent experimental model for tissue regeneration. Furthermore, although the liver consists of various types of cells, hepatocytes account for about $80\%$ of liver weight and about $70\%$ of all liver cells \cite{miya9}. Thus, hepatocytes provide an ideal starting point to study the relation of organ size with number and size of cells.
	
	It has been generally accepted that liver regeneration depends mainly on the proliferation of hepatocytes \cite{miya7, miya10, miya11}. However, there are several reports showing hypertrophy of hepatocytes in the regenerated liver \cite{miya12, miya13, miya14}. Miyaoka et al. \cite{miyaoka1, miyaoka2} performed a series of experiments and found that although a number of studies indicated that almost all hepatocytes proliferate after $70\%$ PH, cellular hypertrophy significantly contributes to liver regeneration as well. They showed that hepatocytes undergo cell division only about $0.7$ times on average in the regeneration from $70\%$ PH, and that at early stages, the regeneration totally depends on the hypertrophy of hepatocytes. In contrast, liver regeneration after $30\%$ PH is  solely due to hypertrophy.
	
	Therefore, liver regeneration process is a perfect scenario to test and calibrate our model. On the other hand, post-hepatectomy liver failure is a serious complication after liver resection and its incidence varies from $1.2$ to $32\%$ \cite{phlf1, phlf2, phlf3, phlf4}. It is defined as functional deterioration of the liver associated with an  increased international normalized ratio (INR) - a measurement of how long it takes blood to form a clot-, and hyperbilirubinemia typically starting after the fifth postoperative day \cite{phlf1}. There are recommendations that post-hepatectomy liver failure could be prevented if the  remnant liver size is above  $20\%$ of its original size in patients with normal liver function, and $30-40\%$ in patients with steatohepatitis or cirrhosis \cite{phlf5, phlf6}. Nevertheless, even with adequate preoperative assessments, post-hepatectomy liver failure is a major contributor to mortality rates of up to $5\%$ after liver resection \cite{phlf7, phlf8}. Various patient-related factors (age, comorbidities such previous chemotherapy, cirrhosis, fibrosis, cholestasis, and steatosis), and surgery-related factors (extent of resection, blood loss, and ischemia reperfusion injury) affect the regenerative capacity of the remnant  liver  \cite{phlf9, phlf10}. Given all these numerous factors, estimating the adequate extent of the hepatectomy, and individual regenerative capacity, remain  significant challenges for clinicians and scientists. It is to be stressed here that the regeneration process is controlled by different transcriptomic signatures depending on the intensity, duration and etiology of liver injury \cite{llovet}. Consequently, different transcripts can modulate the results of the regeneration processes.
	
	In any case, according to the Barcelona Clinic Liver Cancer (BCLC) staging system \cite{C6}, hepatic resection (partial hepatectomy) can be considered as a curative treatment for patients with stage 0-A uninodular hepatocellular carcinoma (HCC) who maintain preserved liver function and without portal hypertension. The prognosis of HCC patients has improved because of advances in radiologic assessment, patient selection, operative techniques, and perioperative care \cite{rec1, rec2}. On the other hand, long-term prognosis of patients with HCC after liver resection is often affected by high tumor recurrence rates that reach $40-70\%$ within $5$ years \cite{C7}. This is an issue that must urgently be addressed, and where we believe that a well calibrated computational model that succeeds on modeling the liver regeneration processes and  HCC evolution, would  complement the clinical trials, once fed-in with specific patient data. Moreover, liver regeneration is the basic element for the maintenance of liver function and size during homeostasis and liver injury (acute and chronic). Understanding the mechanisms of hepatic regeneration, from its cellular origins and signaling mechanisms, is essential to design specific regeneration models. Making these models available in a more “accessible” way that allows for a quicker evaluation of the influence of certain factors on regeneration processes, unlike the ``classical'' models,  would then facilitate the  optimization the different diagnostic strategies (time intervals for hepatocarcinoma screening) and treatments (recurrence rate compared to ablative therapies, immunotherapy, response to sequential therapies). The approach here presented might well constitute a potential tool to evaluate biomarkers, such as circulating tumor cells (determined by liquid biopsy) and their correlation with the rate of tumor growth.
	
	The manuscript is organized as follows: After a brief description of the biological model, our agent-based model and the tumor growth model in section \ref{sec:methods}, we calibrate and validate our model against literature data \cite{miyaoka1, miyaoka2,C1, C2, C3, C4}, for two cases of liver regeneration: $30\%$ PH  and $70\%$ PH (Sections \ref{sec:results;subsec:regeneration;subsubsec:30} and \ref{sec:results;subsec:regeneration;subsubsec:70} respectively). Then we test our model for a $50\%$ PH in section \ref{sec:results;subsec:regeneration;subsubsec:50}. Finally, we have simulated the recurrence of a hepatocellular carcinoma after a PH and present our most significant results in section \ref{sec:results;subsec:carcinoma}. Therein, all our results are commented and analyzed in the light of  experimental (data obtained on mice) and clinical data. Discussion and future prospects are presented in Section \ref{sec:discussion}.

	\section{Materials and methods}
	\label{sec:methods}
	
	\subsection{Biological model}
	\label{sec:biologicalmodel}
	
\begin{figure}[!htb]
		\centering
		\includegraphics[width=0.65\linewidth]{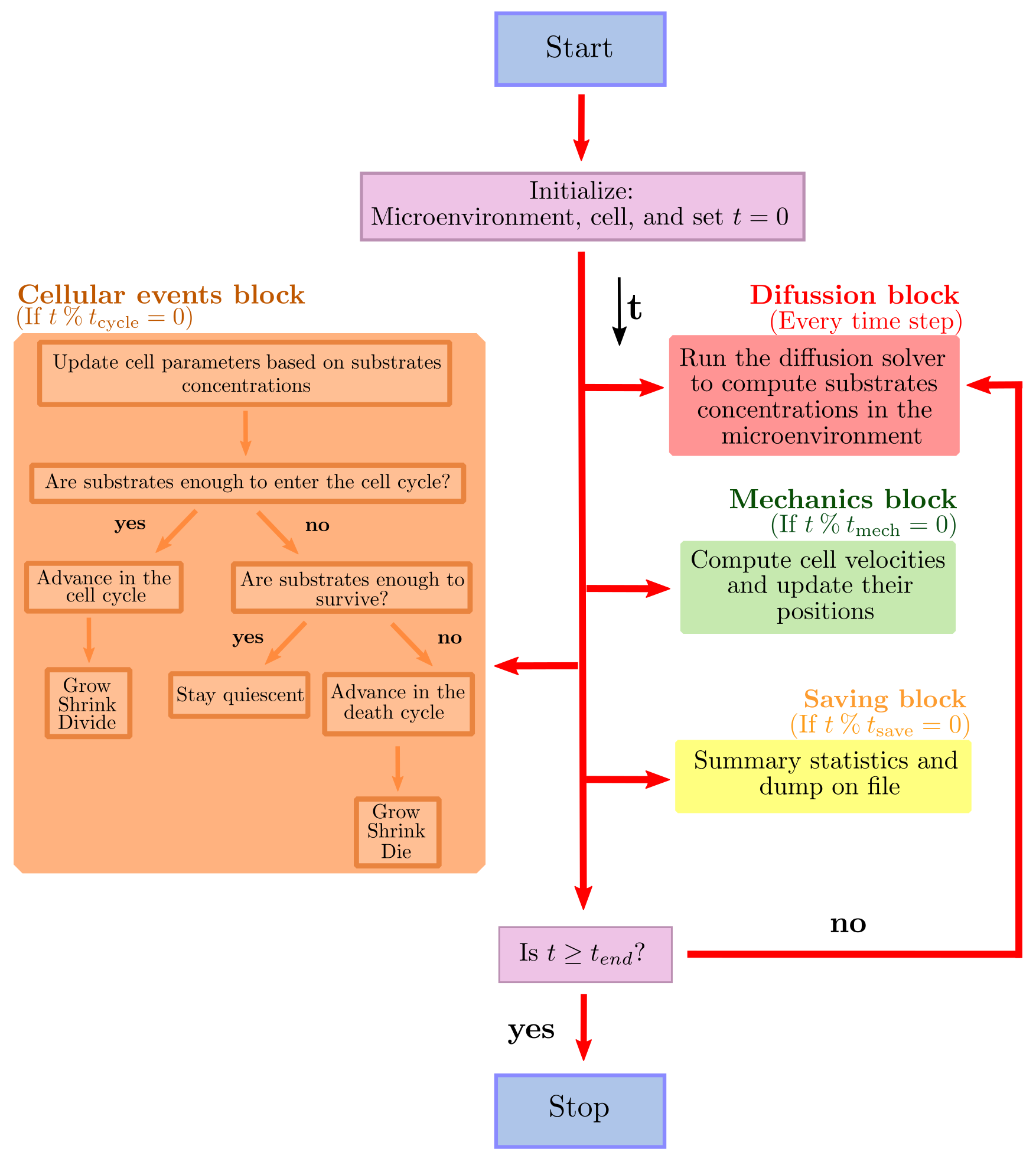}
		\caption{\textit{Schematic representation of the multi-scale liver architecture.} The human liver is divided grossly into four parts or lobes. The four lobes are the right lobe, the left lobe, the caudate lobe, and the quadrate lobe. Seen from the front the liver is divided into two lobes: the right lobe and the left lobe. It is further divided in eight functionally independent segments based in the the Couinaud classification of liver anatomy. At the microscopic (histological) scale, the liver is organized in repetitive functional units called liver lobules, which take the shape of polygonal prisms (typically hexagonal in cross section). Each lobule is mainly constituted by hepatocytes and it is centered on a branch of the hepatic vein called the central vein which is interconnected with the interlobular portal triads: the hepatic artery (red), the portal vein (blue), bile duct (green).}
		\label{higado}
	\end{figure}

	The liver is a highly complex organ, which removes drugs and toxins from the blood. It is characterized by its multi-scale architecture (figure \ref{higado}) which consists of four lobes: the right lobe, the left lobe, the caudate lobe, and the quadrate lobe, which are further divided into eight segments based in the Couinaud system, also known as hepatic segments \cite{boyer}.  Each segment has its own vascular inflow, outflow and biliary drainage. The division of the liver into independent units means that segments can be resected without damaging the remaining segments. Hepatic parenchyma is organized in repetitive functional units called liver lobules. The lobules are roughly hexagonal prisms, and consist of plates of hepatocytes, and sinusoids radiating from a central vein towards an imaginary perimeter of interlobular portal triads. The portal triad is a distinctive component of a lobule, which can be found running along each of the lobule's corners. It consists of the hepatic artery, the portal vein and the common bile duct. The central vein joins the hepatic artery and the portal vein to carry blood out from the liver \cite{ross}. A recent work reported that this particular structures remain during the liver regeneration process \cite{lobulillos}. Even though hepatocytes (liver parenchymal cells) account for about $80\%$ of liver weight and about $70\%$ of all liver cells, the liver also has other type of cells named nonparenchymal cells: endothelial cells, Kupffer cells (macrophages resident in the liver), and biliary-duct cells \cite{miya9}.
	
	One of the main characteristics of the liver is its high regenerative capacity after injury. Even when $70\%$ of its mass is surgically removed, the remnant portion expands to compensate for the lost tissue and functions \cite{miya7,clavien}. Liver resection is the most common liver surgery and consists of removal of liver tissue due to focal lesions, most often malignant tumors and living liver donation \cite{abdelaye}. The multilobular structure of the liver not only allows the surgical resection of a lobe of choice to achieve different degrees by partial hepatectomy but also the resection of lobes does not induce damage to the remaining tissue. The extent of resection is determined by the size and location of the focal lesion and the estimated function of the future liver remnant \cite{nilsson}. Prior to liver resection, surgeons have to assess the patient’s individual risk for postoperative liver dysfunction. In case of malignant tumors, surgeons have to identify the surgical strategy best suited to allow radical oncological removal in order to avoid recurrence, without putting the patient at risk of postoperative liver failure due to excessive removal of liver mass \cite{modelocomputacional, vandam, kan}.
	
	The liver regenerates in a highly organized fashion after surgery \cite{clavien}. The human body responds to partial hepatectomy not by regenerating lost segments but by inducing hyperplasia in the remnant liver \cite{taub,michalopoulos2005,fausto}. The anatomical structures of a liver that has undergone partial hepatectomy are therefore distinctly different from those of the original liver. The process of restoration of liver volume is initiated by the replication of various types of intrahepatic cells, followed by an increase in cell size. Nonparenchymal cells replicate in a delayed fashion. After replication has been completed, growth consisting of an increase in cell size occurs over several additional days. The initiation and synchronization of replication in different types of hepatic cells depend on the extent of the resection, tissue damage, or both. Low-grade tissue damage (\textit{e.g.}, toxic or ischemic injury) or a relatively small resection (removal of less than $30\%$ of the liver) substantially reduces the replication rate, which also appears to be less synchronized than after a large resection (removal of $70\%$ of the liver)\cite{taub,clavien,fausto}.
	
	Since all our results are compared with experimental data obtained from rodent models, it is important to mention that the process of hepatic regeneration in rodents and humans is similar, and the results obtained from rodents could be applicable to the human liver \cite{fausto2}. Moreover, the rat liver architecture can be compared with the human liver segmentation defined by Couinaud \cite{kouge}.
	
	\subsection{Computational model}
	\label{sec:modelocomp}
	
	Our model is implemented resorting to an object oriented programming model, and to that aim we have used C++11 language. Simulation CPU time depends on model parameters such as domain (lattice) size, cell number and simulation length (in time); a typical simulation run takes approximately $6$ h on a single core of an Intel i7-10510U CPU. Model visualization is performed with Ovito \cite{ovito}, Paraview \cite{paraview} and Matplotlib \cite{matplotlib}.
	
	In order to reduce the computational burden of the simulations, an abstraction process was necessary to go from the biological to the computational model. First, we disregard the explicit liver geometry, instead we have chosen a reduced spherical model. This simplification is possible because, as it was mentioned in section \ref{sec:biologicalmodel}, after a PH the liver does not regenerate the lost segments, \textit{i.e.} does not recover the original shape. Instead it just recovers its original volume by hyperplasia of the remaining lobes. Subsequently, we rather focus our attention on the liver parenchyma instead of liver lobes. In our computational model, hepatic lobules are hexagonal prisms delimited by an imaginary perimeter of interlobular portal triads. The central vein that carries the blood out from the liver as well as the liver sinusoids are not explicitly modeled as the portal triads are. We rather model this behavior by tuning the effective oxygen difussion and decay constants. A discussion on the implications of these simplifications on the results will be presented in section \ref{sec:discussion}.
	
\begin{figure}[!htb]
		\centering
		\includegraphics[width=0.8\linewidth]{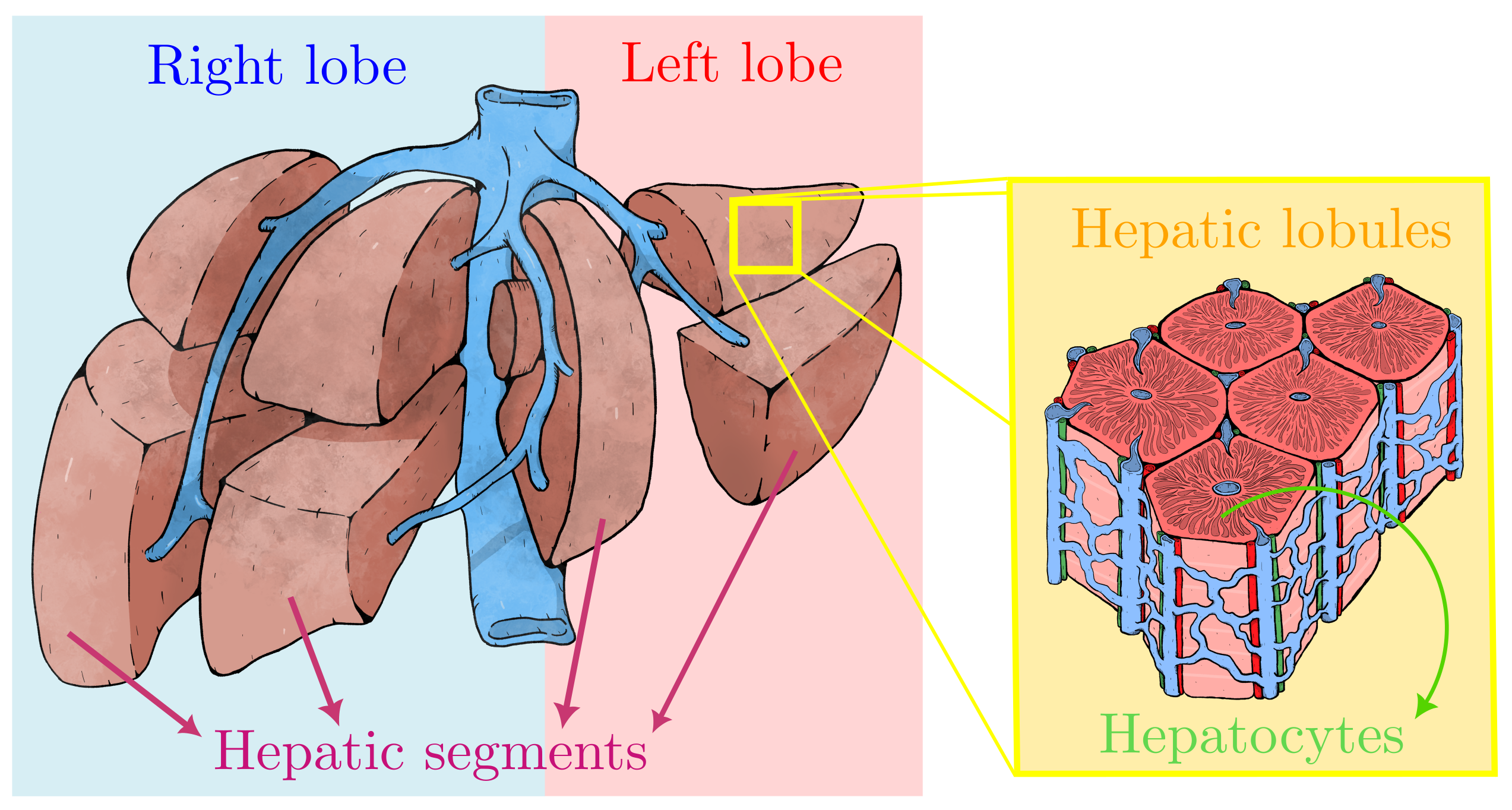}
		\caption{\textit{Main loop flow diagram.} Blue box represents the start of the program. Red box represent the diffusion processes. Green box and orange box describe the cell mechanics and cycling processes respectively. Finally, yellow boxes represent the data saving process. After initializing the microenvironment, the cells, and the current simulation time $t = 0$, our model tracks (internally) $t_{\text{mech}}$ (the next time at which cell mechanics methods are run), $t_{\text{cycle}}$ (the next time at which cell processes are run), and $t_{\text{save}}$ (the simulation data output time), with output frequency $\Delta t_{\text{save}}$. $\%$ represents the modulo operation.}
		\label{diagrama}
	\end{figure}

	In the following subsections we will describe the methods implemented to model diffusion and cellular mechanics, as well as the mathematical models to predict tissue growth kinetics. For further details, we refer the reader to the supplementary material. A schematic representation of the inner workings of our model is depicted in Fig. \ref{diagrama}. The key elements of the model are described in what follows.
	
	\subsubsection{Diffusion solver}
	\label{sec:methods;subsec:diffusion}
	We model the diffusion of chemical substrates in the tumor microenvironment as a vector of reaction-diffusion partial differential equations for a vector of chemical substrates, $\rho$. It is discretized over a Cartesian mesh for computational convenience, in such a way that each voxel (volumetric pixel) stores a vector of chemical substrates. Each substrate diffuses and decays, and can be secreted or uptaken by individual cells at their specific positions.
	
	We use a first order, implicit (and stable) operator splitting, allowing us to create separate, optimized solvers for the diffusion-decay, cell-based source/sinks \cite{marchuk}. The diffusion-decay terms are solved using the finite volume method \cite{fvm}, further accelerated by an additional first-order spatial splitting  in the $x-$, $y-$ and $z-$directions via the locally one-dimensional method (LOD) \cite{marchuk, lod}. For each dimension, we solve the resulting tridiagonal linear systems with the Thomas algorithm \cite{thomas}.
	
	We also implement the so-called Dirichlet nodes, so that substrate values at any voxel within the simulation domain can be overwritten  to turn the voxel into a continuous source of substrates. This is particularly useful to model the effect of  blood vessels, or when applying Dirichlet boundary conditions.
	
	For computational efficiency we use thread parallelization to  relevant loops, e.g. in many instances of the Thomas solver when solving $x-$diffusion across multiple strips of the solution domain. This methods were already implemented and successfully tested by Ghaffarizade et al. \cite{biofvm}, therefore, we have validated the numerical accuracy of our solver by comparing our results with those found in Reference \cite{biofvm}. For further details, please refer to the supplementary material (S1 Text)

	\subsubsection{Cell agents}
	\label{sec:methods;subsec:cell}
	
	Since we are implementing an agent-based model programmed in the context of an object oriented approach, each cell is an agent implemented as a software object that acts independently. Like most classes in software it has \textit{attributes}, \textit{i.e.} its own internal variables that each specific agent is allowed to manipulate on its own. It also has \textit{methods}, which are functions that act upon the attributes. In the context of cell biology, relevant  attributes might be: position, size, cell state regarding  the cell cycle, etc. The cell cycle is an object with the aforementioned attributes. The cell class have  methods that represent cellular processes such as, death, growth and volume change, and are coordinated by the cell cycle object.
	
	One of the main features of our model is that cells are \textit{off-lattice}. Consequently, they are not confined to a particular lattice or spatial arrangement, they move in a continuous fashion through all space positions, and therefore underlying possible artifacts associated with the chosen lattice structure and spacing are removed.
	
	Based on previous cell-based models \cite{physicell, episim, chaste}, we have modeled the cell behavior as follows:

	\paragraph{Cell cycle:}
	\label{sec:methods;subsec:cell;subsubsec:cycle}
	We model the cell cycle as a directed graph, in which the nodes represent the phases and the edges the transition rates between them. These transition rates can take stochastic or constant values. Moreover, any of the cell cycle time scales can be adjusted at the beginning of the simulation to match different types of growth and they can also be adjusted at any time on an individual cell in order to reflect its microenvironment influences.
	
	Our model allows to implement different types of cell cycles based on different parameters. Following Miyaoka and coworkers \cite{miyaoka1,miyaoka2}, we can base our cell cycle on a tracking of the expression of protein Ki-$67$ \bibnote{Ki-$67$ protein expresses in the $S$, $G2$, and $M$ phases \cite{ki675}, and to a lesser extent in the $G1$ phase \cite{ki6750}. It is seen in post-mitotic daughter cells \cite{ki672}, but it is not produced in these cells \cite{ki675}. Instead, any remaining Ki-$67$ protein in post-mitotic cells is degraded quickly, with a half-life of $60-90$ minutes \cite{ki675}. We can base our cell cycle by setting the phase ``Ki-$67$+ pre-mitotic'' to be the combined duration of $S$, $G2$ , and $M$ phases, which are relatively fixed compared to the duration of $G0/G1$ \cite{ki6732, ki6733}. We can set the second phase ``Ki-$67$+ post-mitotic'' at the $G1/G0$ phase, to be on the order of two Ki-$67$ half-lives. Then, the remaining $G1/G0$ phase will be the ``Ki-$67$-'' phase. For further details, please refer to the supplementary material S1 Text.}, involved in cell proliferation. This tracking  is performed thanks to a nuclear stain of Ki-$67$. 
	
	Finally, since at cell scale death is not an instantaneous event but a process, we model death cycles such as Necrosis and Apoptosis as we did with cell cycles, by using directed graphs, death is part of the cell cycle after all.  The entry to the death cycles depends on the resources, for example oxygen, drugs, therapies, etc.
	
	\paragraph{Cell volume:}
	To model cell volume variation, each cell tracks $V$ (total volume), $V_F$ (total fluid volume), $V_S$ (total solid volume), $V_{NS}$ (nuclear solid volume), $V_{CS}$ (cytoplasmic solid volume), $V_N$ (total nuclear volume), and $V_C$ (total cytoplasmic volume). Key parameters include nuclear solid, cytoplasmic solid, and fluid rate change parameters ($r_N$, $r_C$, and $r_F$), the cell’s ``target" fluid fraction $f_F$, target solid volume $V_{NS}^{*}$, and target cytoplasmic to nuclear volume ratio $f_{CN}$. For each cell, these volumes are modeled with a system of ordinary differential equations that allow cells to grow or shrink towards a target volume. These parameters are updated as the cell progresses through its current cycle or death cycle. 
	
	\paragraph{Cell mechanics:}
	In our model, as in \cite{physicell}, we use the piecewise polynomial force which is constructed as the sum of a positive adhesive and a negative repulsive polynomial force contributions. It is important to note that  repulsive forces are really an elastic resistance to deformation. To compute these forces we use adhesive and repulsive interaction potentials functions.
	
	We assume that three types of forces act upon each cell. First we have a drag force, which represents dissipative drag-like forces such as fluid drag and cell-extra cellular matrix adhesion. We then have neighboring cell mechanical forces. In the simplest case these involve repulsive forces due to limited cell compressibility, but they usually also include cell-cell adhesion. We use interaction potentials that depend upon each cell's size, maximum adhesion distance, adhesion and repulsion parameters and distance to other cells. Finally, the third force acting on the cells is the cell-basement membrane interaction.
	
	Since the cell microenvironment has a very low Reynolds number (the Reynolds number describes the ratio of inertial to viscous forces) \cite{reynolds}, inertial effects such as acceleration are neglected. This is commonly known as the inertialess assumption, and implies  that forces equilibrate at relatively fast time scales in contrast to the processes involved in  cell cycling, death cycling, and volume variation.
	
	We finally use the Adams-Bashforth method \cite{griffiths} to solve the mechanics equation to enhance computational efficiency.
	
	We refer the reader to the supplementary material (S1 Text) for further details on the implementation of the piecewise polynomial force model.
	
	\paragraph{Cell secretion and uptake:}
	\label{sec:methods;subsec:cell;subsubsec:secretion}
	
	This is one of the most important data structures of the cell because it links the cell with its microenvironment. We solve a vector of partial differential equations, which reduce to the addition of  a cellular secretion/uptake term to the diffusion equation described in section \ref{sec:methods;subsec:diffusion}.
	
	This is very important since most of the cellular processes depend on the substrates that diffuse in the microenvironment. For example, it is well accepted that after a partial hepatectomy, the liver undergoes to cytokine- and growth factor- mediated regeneration processes \cite{markers}. However, most of the mechanisms of initiating and promoting regenerative responses as well as the termination of liver regeneration remain unknown \cite{miyaoka1, miyaoka2}. In this work cellular proliferation is assumed to be controlled by a growth factor that diffuses through the microenvironment. This growth factor is only considered as an abstract parameter which encompasses all the underlying molecular mechanisms involved in the liver regeneration process. The cell cycle entry rate is proportional to this factor in the following way:
	
	\begin{equation}\label{o2prolif}
		r =\frac{1}{t_{K^{-}}}\text{max} \left\{ \left( \frac{GF - GF_{\text{prol}}}{GF^{*} - GF_{\text{prol}}} \right), 0 \right\}
	\end{equation}
	
	\noindent where $t_{K^{-}}$ is the mean time each cell  spends in the non-proliferative phase (see section \ref{sec:methods;subsec:cell;subsubsec:cycle}), which can be experimentally monitored using the Ki-$67$ cellular marker. $GF$ is the current growth factor concentration value in the cell's voxel, $GF_{\text{prol}}$ is the proliferation threshold, \textit{i.e.} the growth factor value below which the proliferation ceases and $GF^{*}$ is the proliferation saturation value, above which the proliferation rate is maximized. Therefore, based on the growth factor concentration, the hepatocyte will enter either the hypertrophy phase or the proliferation phase. A similar approach can be done based on the oxygen concentration, however, instead of influence on the decision about whether or not to proliferate, oxygen will accelerate the phase entry. Please refer to the supplementary material (S1 Text) for further information about how cell cycles and transition rates will be modified based on chemical substrates concentrations.
	
	\subsection{Growth estimations}
	
	To complement our model in the prediction of tumor growth and/or tissue regeneration we use a mathematical model known as the \textit{Gompertz model} \cite{gompertz7, gompertz8, gompertz9, gompertz}. This model assumes initial exponential growth, but as the tumor grows the volume-doubling time increases due to lack of nutrients and subsequent cell death, by which the growth rate shifts towards linear regime,  finally reaching a plateau  \cite{gompertz7}. This is given by:
	
	\begin{equation}\label{gompertz}
		V(t)=K\exp{\left[\log{\left(\frac{V_{0}}{K}\right)} \exp{\left(-\alpha t\right)}\right]}
	\end{equation}
	
	\noindent where the parameter $K$ is the carrying capacity of the tissue, which is the highest possible volume, $V_{0}$ is the initial volume of the tissue and $\alpha$ is the specific growth rate \cite{sgr,sgr2} which is defined by
	
	\begin{equation}\label{sgr}
		\alpha = \frac{\log{\left(\frac{V_{2}}{V_{1}} \right)}}{t_{2} - t_{1}}.
	\end{equation}
	
	Here $V_{1}$ and $V_{2}$ are tumor volumes at the measure times $t_{1}$ and $t_{2}$ respectively. This parameter $\alpha$ determines how fast the tumor reaches the carrying capacity $K$ and its measured in ($\%/\text{days}$).

	\section{Results}
	\label{sec:results}
	
\begin{figure}[!htb]
		\centering
		\includegraphics[width=0.8\linewidth]{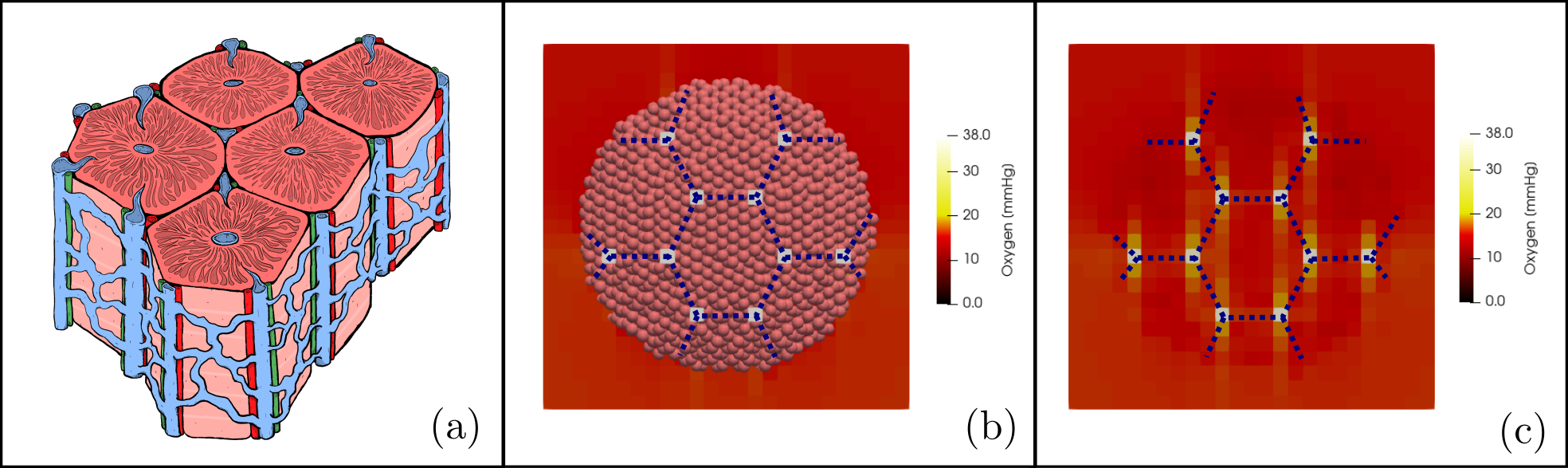}
		\caption{\textit{Liver blood vessels architecture.} \textbf{(a)} Schematic representation of the hepatic lobules. They consist of plates of hepatocytes, and sinusoids radiating from a central vein interconnected with the interlobular portal triads: the hepatic artery (red), the portal vein (blue), and the common bile duct (green). \textbf{(b)} computational model of the liver parenchyma in which we can observe the blood vessel architecture. \textbf{(c)} Heat map of the oxygen diffusion in the liver microenvironment. Blue dotted lines were drawn just as a guide to the eye.}
		\label{vasos}
	\end{figure}

	We first attempted to define a baseline scenario of the liver physiology. As it was mentioned before, we focus our attention in the liver parenchyma which, in our model, is made up of hepatic lobules which are hexagonal prisms delimited by an imaginary perimeter of interlobular portal triads. We idealized this dynamic vasculature architecture (figure \ref{vasos}a) by using the Dirichlet nodes as shown in figures \ref{vasos}b and \ref{vasos}c, and defining the distance between triads on micrograph analysis \cite{ross}. Figure \ref{vasos}b shows a transversal cut of our simulated liver in which we can observe the hepatic lobules architecture. Blue dotted lines were drawn just as a guide to the eye. Pink spheres represent the hepatocytes and the white squares represent the portal triads that oxygenate the tissue. Figure \ref{vasos}c shows the heat map of the oxygen diffusion in the liver micoenvironment. The oxygen diffuses from the portal triads (Dirichlet nodes), there is no diffusion from the boundaries of the simulation box. 
	
	The grid we used to simulate liver regeneration is $1 \; \text{mm}^3$ in total size, orders of magnitude smaller than an actual liver. While the model does not place limitations on the liver size, the implementation is obviously constrained by the size of the computer memory. However, this limitation can be to some extent circumvented by considering the sample region as representative of a significant sub-region of the liver. Obviously size effects must be investigated in further detail to the extent that new algorithmic and hardware improvements become available (e.g. by exploiting massively parallel CUDA \cite{cuda} cores and/or Tensor cores in new GPU/TPU architectures).
	
	All the parameters values that were used in the following examples, and the corresponding sources are listed in section $4$ of the supplementary material (S1 Text).
	
	\subsection{Liver regeneration}
	\label{sec:results;subsec:regeneration}
	
\begin{figure}[!htb]
		\centering
		\includegraphics[width=0.8\linewidth]{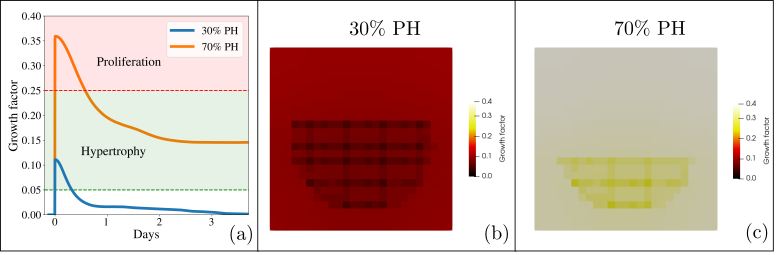}
		\caption{\textit{Growth Factor.} \textbf(a) Growth factor concentration in the liver microenvironment after a $30\%$ PH (blue line) and a $70\%$ PH (orange line), and the response it will cause in the hepatocytes. \textbf(b) and \textbf(c) Heat map of the growth factor concentration after $30\%$ PH and a $70\%$ PH respectively. The concentration is greater in the $70\%$ PH.}
		\label{factor}
	\end{figure}

	Our first experiments aim at assessing the ability of our model to describe the dynamics of the liver regeneration process. We performed the \textit{in silico} version of the experiments of Miyaoka et al. \cite{miyaoka1, miyaoka2},  with the novelty that hepatocytes secrete and are sensitive to a growth factor that diffuses through the microenvironment creating a gradient. We assume that when the liver is intact, the microenvironment is in homeostasis and, although they are metabolically active, hepatocytes remain dormant in the cell cycle. When a partial hepatectomy occurs, each cell secretes a constant amount of growth factor depending on the extent of  the injury.  They will undergo either hypertrophy or proliferation as regulated by the concentration of the growth factor. This is illustrated in figure \ref{factor}. In panel (a), we can see a peak of growth factor when the partial hepatectomy occurs. A $30\%$ PH does not generate enough growth factor to make hepatocytes proliferate (figure \ref{factor}b). On the contrary, a $70\%$ PH not only produces enough growth factor to make the hepatocytes proliferate, they also undergo an hypertrophy process (figure \ref{factor}c). Finally, as the liver regenerates, this growth factor decreases.
	
	\subsubsection{30\% partial hepatectomy}
	\label{sec:results;subsec:regeneration;subsubsec:30}
	
\begin{figure}[!htb]
		\centering
		\includegraphics[width=0.75\linewidth]{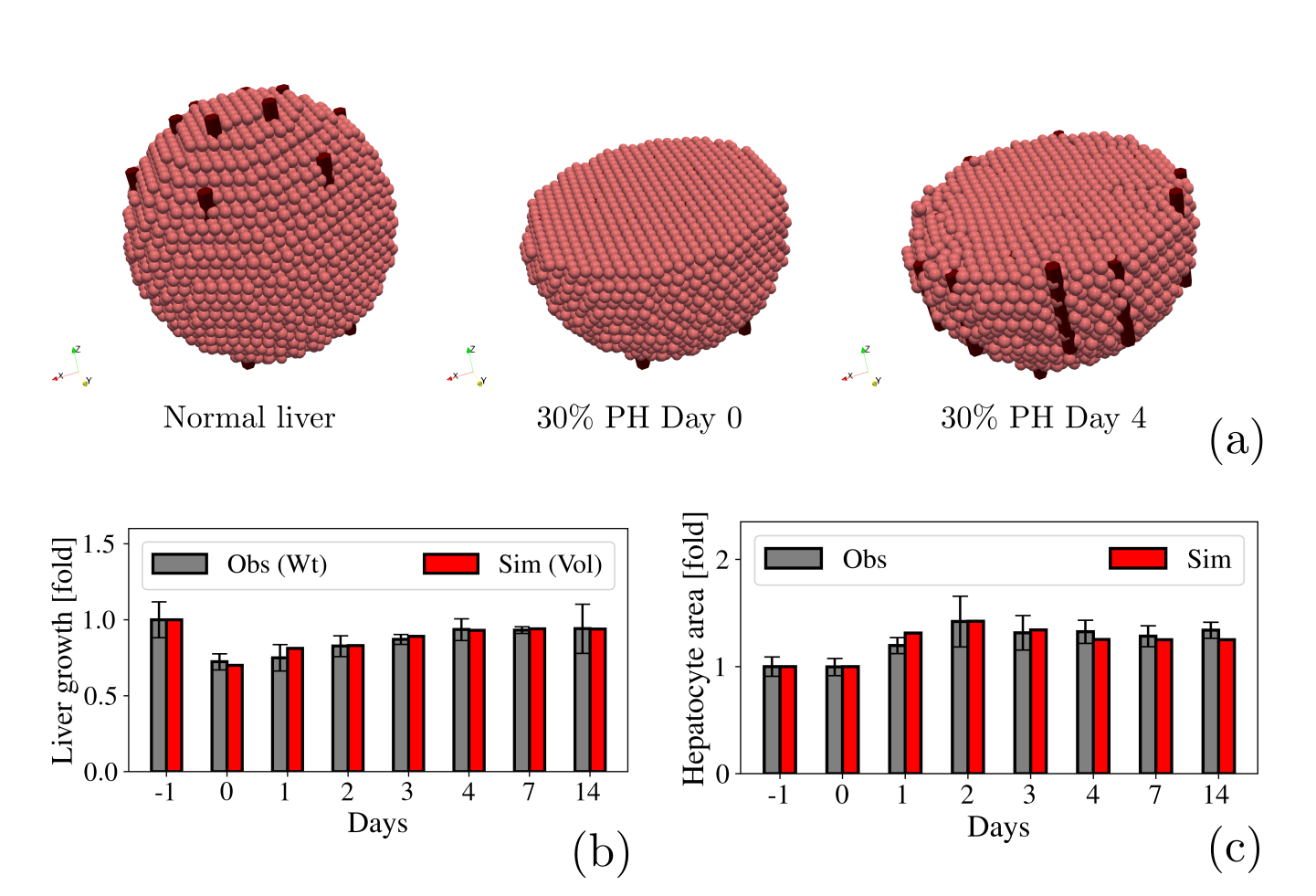}
		\caption{\textit{$30\%$ PH} \textbf{(a)} Qualitative representation of the liver regeneration process after a $30\%$ PH. \textbf{(b)} Fold-increase in the liver size. Observational data is shown in gray and represents the weight of the liver measured by Miyaoka et. al \cite{miyaoka1, miyaoka2}. Simulated data is shown in red and represents the liver volume. \textbf{(c)} Quantification of the hepatocytes area during liver regeneration. }
		\label{30ph}
	\end{figure}

	We first analyzed liver regeneration after a 30\% PH. We have set the hepatocytes cell cycle time $33.6$ hr as stated in ref. \cite{hepatociclo} but as  mentioned in section \ref{sec:methods;subsec:cell;subsubsec:secretion}, this could  vary depending on the oxygenation of the tissue and the growth factor concentration. Figures \ref{30ph}a and \ref{30ph}b illustrate the regeneration process in a qualitative and a quantitative manner respectively. We found that liver volume increased from $1$ day after the PH and reached a plateau of $0.93$-fold of the liver initial volume, by $4$ days. Quantification of the hepatocytes reveals that there was no proliferation but hypertrophy. As shown in figure \ref{30ph}c, hepatocytes reached the largest size  $2$ days after PH, increasing by a factor $1.6$, and then gradually decrease their size only by a $1.4$ factor, thus remaining larger than before the PH. This amounts to a $1.5$-factor increase in volume, suggesting that cellular hypertrophy alone compensates the lost tissue.
	
	Figures \ref{30ph}b and \ref{30ph}c show our results (red bars) in agreement with those collected from the experiments of  Miyaoka et al. \cite{miyaoka1, miyaoka2} (gray bars). It is important to mention that since our model does not account for explicit changes is liver's weight, we have analyzed its volume instead.
	
	An animation of the liver regeneration process after a $30\%$ partial hepatectomy can be seen in the Supplementary Material S1 Video.
	
	\subsubsection{70\% partial hepatectomy}
	\label{sec:results;subsec:regeneration;subsubsec:70}
	
\begin{figure}[!htb]
		\centering
		\includegraphics[width=0.95\linewidth]{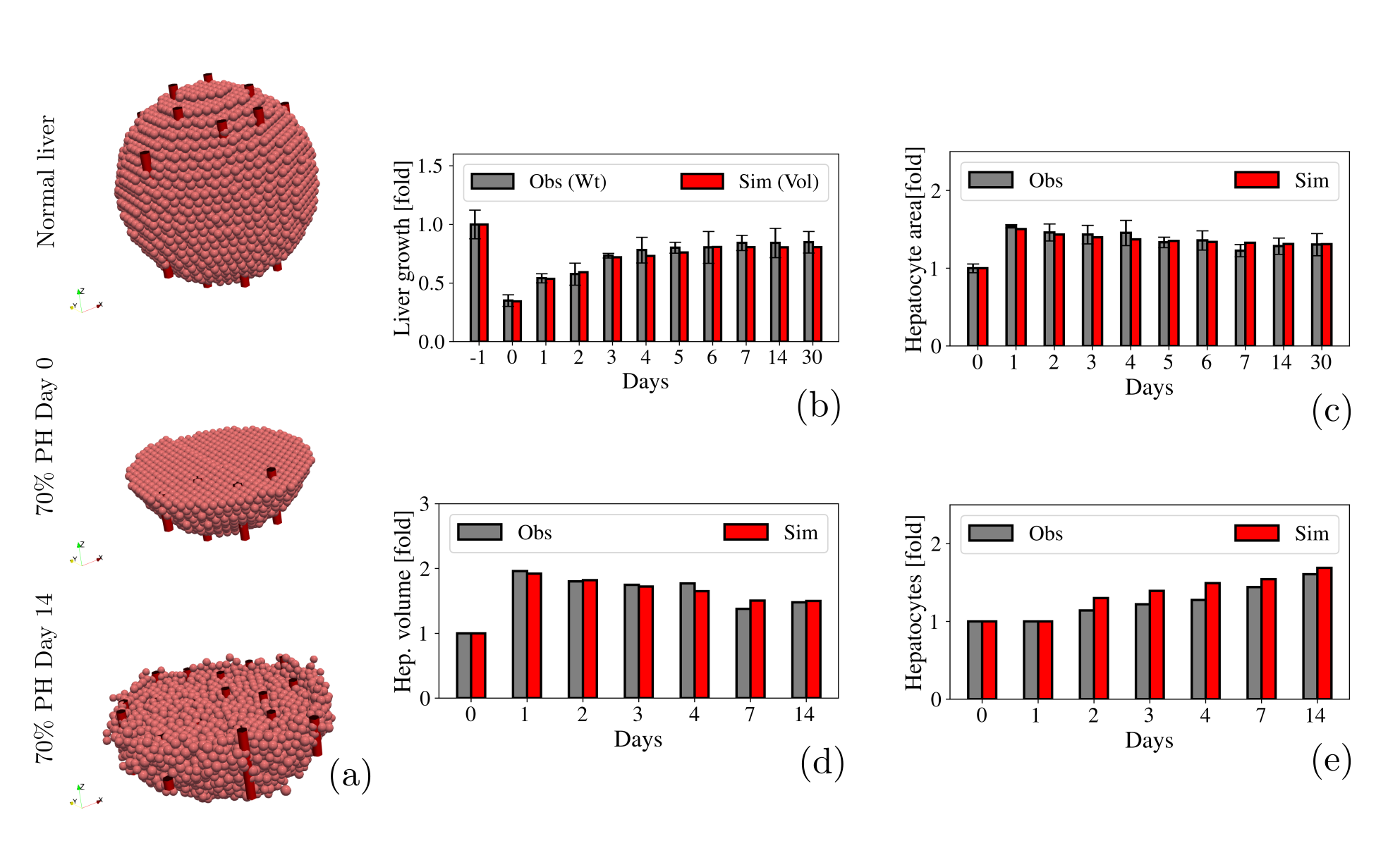}
		\caption{\textit{$70\%$ PH} \textbf{(a)} Qualitative representation of the liver regeneration process after a $70\%$ PH. \textbf{(b)} Fold-increase in the liver size. Observational data is shown in gray and represents the weight of the liver measured by Miyaoka et. al \cite{miyaoka1, miyaoka2}. Simulated data is shown in red and represents the liver volume. \textbf{(c)} Quantification of the hepatocytes area during liver regeneration. \textbf{(d)} Quantification of the hepatocytes volume during liver regeneration. \textbf{(e)} Fold-increase in the number of hepatocytes during liver regeneration.}
		\label{70ph}
	\end{figure}

	After a $70\%$ PH, we found that liver volume increased from day $1$, reaching a plateau with a total increase of $0.72$-fold of the liver initial volume in day $7$. Figures \ref{70ph}a and \ref{70ph}b illustrate this process in a qualitative and a quantitative fashion respectively. Hepatocytes entered the cell cycle  $2$ days after the PH, as shown in figure \ref{70ph}e. Although there were a few active proliferating hepatocytes on day $1$, liver volume had increased considerably by that time. Then we measured the area of the hepatocytes (figure \ref{70ph}c) and found that hepatocytes increased their volume by $2.0$-fold the first day and gradually decrease by $1.5$-fold the $14^{th}$ day after the PH (figure \ref{70ph}d). These results are in agreement with those obtained by Miyaoka et al. \cite{miyaoka1, miyaoka2}, and indicate that proliferation of hepatocytes alone could not account for the recovery of liver mass after a $70\%$ PH.
	
	The animation of the liver regeneration process after a $70\%$ partial hepatectomy, can be seen in the S2 Video.
	
	Although the exact mechanisms underlying liver regeneration have not yet been fully characterized, studies have shown that after $70\%$ PH many of the upregulated growth factors in a regenerating liver are known for their angiogenic properties in vivo. For instance, vascular endothelial growth factor (VEGF) is upregulated after PH \cite{vs15, vs16, vs17}. It is a major pro-angiogenic factor \cite{vs18} and is thought to improve sinusoid reconstruction during the liver regeneration process \cite{vs19}. In the processes of angiogenesis associated with tissue regeneration, two phases are described; an induction phase and another proliferative angiogenesis phase. In post-hepatectomy liver regeneration models, the first is calculated during the first $4$ days and the second between days $4-12$ \cite{C5}.
	
\begin{figure}[!htb]
		\centering
		\includegraphics[width=\linewidth]{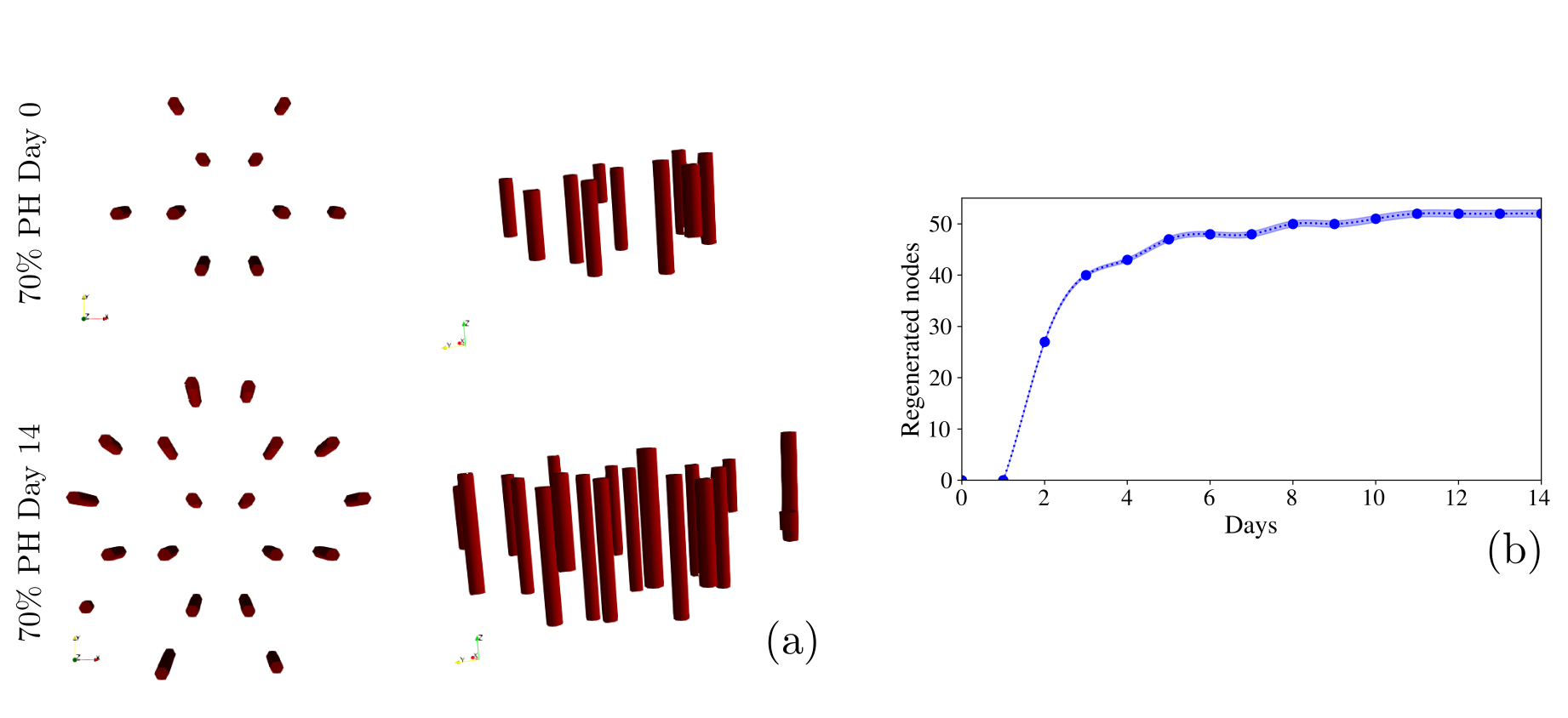}
		\caption{\textit{Blood vessels regeneration} \textbf{(a)} Schematic representation of blood vessels regeneration after a $70\%$ PH. Left panels show an upper view while right panels show a side view. Blood vessels regenerate by keeping the hepatic lobules structure. \textbf{(b)} Quantification of the blood vessels regeneration by counting the Dirichlet nodes added during the regeneration process. It shows a significant increase during the first $3$ days until it finally reaches a plateau. Shaded region represent the standard deviation of $40$ simulations}
		\label{nodos}
	\end{figure}

	We have quantified blood vessel formation by counting the Dirichlet nodes added during the liver regeneration process. Figures \ref{nodos}a and \ref{nodos}b represent this process in a qualitative and a quantitative way, respectively. On one hand we found that during the regeneration process, blood vessels keep the hepatic lobules architecture (figure \ref{nodos}a). To model this behavior, we have labeled each voxel according to its potential to become a blood vessel, inspired by the hepatic lobule architecture. During the simulation, if the voxel is tagged as a potential blood vessel and contains five or more cells, it turns into a Dirichlet node, that will provide oxygen to the cell microenvironment. As shown in figure \ref{nodos}b, we found that the number of Dirichlet nodes increased significantly during the first $3$ days following PH until they finally reach a plateau. These results are in agreement with ref. \cite{vs2}, in which it was found that the micro-vasculature density increased significantly during the first $3$ days in mice subjected to PH. This proves that our model has the ability to simulate cell population behavior, and subsequently it also can predict blood vessel formation.
	
	\subsubsection{Test of a 50\% partial hepatectomy}
	\label{sec:results;subsec:regeneration;subsubsec:50}
	
	Although $70\%$ PH is the most studied instance of liver regeneration \cite{taub,fausto3}, resection of approximately half volume of the donor liver is more common in the case of living donor liver transplantation setting. \cite{clavien}. Moreover, as it was mentioned before, resection must be more conservative in the presence of underlying liver diseases or in elderly patients (e.g., $\geq 70$ years of age) \cite{clavien}. Major ($>50\%$) hepatectomy in the presence of cirrhosis or steatosis significantly increased morbidity \cite{macromac}. With the presence of liver steatosis, $30\%$ or more of the remnant liver should remain in order to maintain viability. Furthermore, studies have revealed an increased benefit of ischemic preconditioning in patients with hepatic steatosis who had lower resected liver volume ($<50\%$) \cite{clavien2}, and extensive resections are generally not recommended for patients with cirrhosis \cite{ray}. Hemihepatectomy (\textit{i.e.} $50\%$ PH) has now been successfully and frequently used for surgical removal of liver associate tumors and cancers \cite{haglund,paperreferee}. However, to the best of our knowledge, a study of liver regeneration dynamics after a $50\%$ PH, such as the one presented in \cite{miyaoka1, miyaoka2} for $30\%$ and $70\%$ PH, has not been reported yet. Using the calibration of the previously studied PH, we have applied our model to study the liver regeneration process after a $50\%$ PH.
	
\begin{figure}[!htb]
		\centering
		\includegraphics[width=0.85\linewidth]{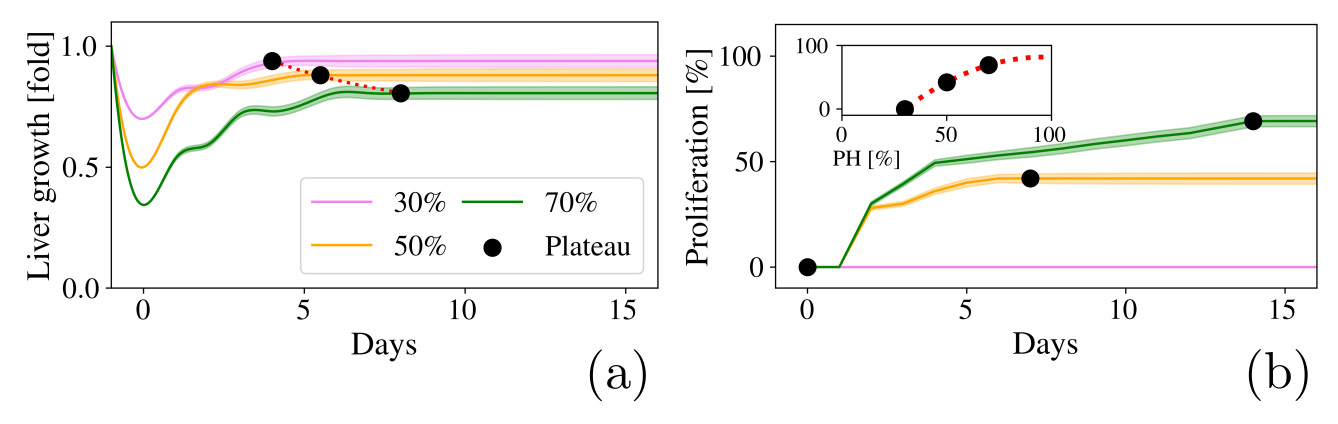}
		\caption{\textit{Predictions of a $50\%$ PH compared to the outcomes of $30\%$ and $70\%$ PH} \textbf{(a)} Fold-increase in the liver volume after $30\%$ (violet), $50\%$ (orange) and $70\%$ (green) PH. Black dots represent the time at which the volume increase reaches a plateau and the red dotted line is a polynomial fitting. \textbf{(b)} Percentage of hepatocytes that proliferates during liver regeneration. Black dots represent the time at which the proliferation increase reaches a plateau. \textbf{Inset:} Proliferation percentage in terms of the PH. The red dotted line represents a polynomial fitting. Shaded regions represent the standard deviations of $40$ simulations}
		\label{50PH}
	\end{figure}

	We found that the liver volume increased from day $1$ until it reaches a plateau with a total increase of $0.86$-fold of the liver initial volume, within $5.5$ days. Yoshioka et al. \cite{50ph} reported that $3$ days after a $50\%$ PH, the remnant liver weight reached $0.72$-fold $\pm 0.05$-fold of its original calculated weight, which is similar to the $0.78$-fold that our model predicted for the third day after the PH. As shown in figure \ref{50PH}a, we can apply a polynomial fitting that matches our simulations and estimates the outcomes of the hepatectomies. Figure \ref{50PH}b shows that, similar to the $70\%$ PH, the hepatocytes not only undergo hypertrophy, they also proliferate. We found that hepatocytes undergo cell division only about $0.42$ times on average and reaches a plateau within $7$ days. If we plot the percentage of hepatocytes that proliferates in terms of the the amount of liver removed, as shown in the inset of figure \ref{50PH}b, we can we can apply a polynomial fitting to estimate the outcomes of different PH.
	
	It is important to mention that the model is calibrated based on an normal hepatic parenchyma, i.e. without liver cirrhosis, with preserved liver function and no signs of portal hypertension, therefore its predictions correspond to that specific case. However it could be adjusted to model the presence of underlying liver diseases.
	
	\subsection{Recurrence of hepatocellular carcinoma}
	\label{sec:results;subsec:carcinoma}
	With the scenario of the liver regeneration working properly, we are now interested in study tumor recurrence. The need for extended liver resection is increasing due to the growing incidence of liver tumors in aging societies \cite{modelocomputacional}, however, the resected volume not only depends on the tumor volume itself but also on the patient’s liver overall health. Prior the resection, surgeons not only have to assess tumor resection to avoid recurrence but also the patient’s individual risk for postoperative liver dysfunction. It is well known that planning for a safe resection of a liver tumor with a large future liver remnant reduces the risk for postoperative liver failure but increases the risk of tumor recurrence. In contrast, planning for an oncologic radical surgery requires to remove a safety margin. Extending the safety margin in case of a centrally located tumor leads to a substantially extended resection leaving a rather small future liver remnant behind, which increases the risk of postoperative liver failure \cite{modelocomputacional}. In a patient with uninodular hepatocarcinoma or up to $3$ nodules smaller than $3$ cm each without macrovascular infiltration, without distant metastasis (BCLC Stage 0-A), with preserved liver function and without portal hypertension, a liver resection of up to $60-70\%$ is a feasible scenario considered in the BCLC classification \cite{C6}, as long as it presents a normal hepatic parenchyma. Preexisting liver disease such as steatosis increases the risk for postoperative liver failure and might therefore call for a smaller PH compared to livers without preexisting diseases, however, the study of liver and hepatocellular carcinoma dynamics with preexisting diseases exceeds the scope of this paper.
	
	Following the previous case in which we have considered a normal hepatic parenchyma, we have seeded a remaining tumor clone in order to model a recurrence of a hepatocellular carcinoma after a extended resection, \textit{i.e.} a $70\%$ PH. As proof of concept, we have assumed that preserved liver function without portal hypertension could approximate the behavior of a liver with preserved or normal parenchyma.
	
\begin{figure}[!htb]
		\centering
		\includegraphics[width=0.7\linewidth]{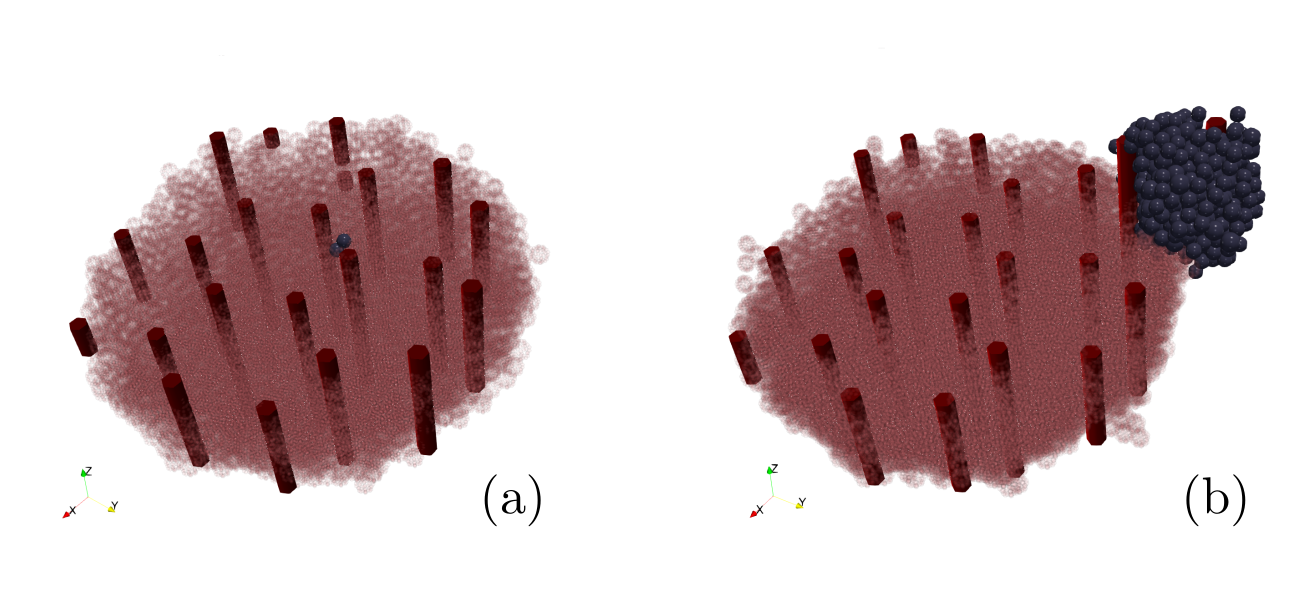}
		\caption{\textit{Random initial position.} Qualitative representation of the liver regeneration with HCC recurrence, $30$ days after the PH. Hepatocytes are represented by transparent pink cells to have a better view of the blood vessels (red tubes) and the tumor (blue cells) growth. \textbf{(a)} The slowest tumor growth happens when the residual tumor clone was seeded in the center of the liver surface. \textbf{(b)} The fastest tumor growth happens when the residual tumor clone was seeded in the periphery of the liver.}
		\label{position}
	\end{figure}

	Cancer cell cycle has a duration of $38.6$ hr \cite{hepatociclo} but the rate of cycle entry scales proportionally to oxygen concentration. 
	As a first step, we have randomly seeded the tumor clone over the liver surface in order to test if its initial location would change the simulation outcomes. We have performed $40$ simulations and obtained a $59\%$ standard deviation of the tumor final size. Figure \ref{position} shows the smallest (a) and the biggest (b) tumors, $30$ days after the PH. Hepatocytes are drawn transparent to have a better view of the blood vessels (red tubes) and the tumor (blue cells) growth. We observe that the tumor volume vary depending on the location of the residual tumor clone, whether it is in the periphery or the center of the liver surface. This might be due to the process of blood vessels generation implemented in our model (sec. \ref{sec:results;subsec:regeneration;subsubsec:70}). When the tumor clone is located in the center of the liver surrounded by hepatocytes, it will have to compete for oxygen with the surrounding cells (\textit{i.e.} share the preexisting blood vessels), reducing the cell cycle entry. On the contrary, when the tumor clone is located in the periphery, it will grow outwards. That means that the tumor cell not only has more oxygen for itself due to the liver blood vessels, but as the tumor gets bigger, it generates its own blood vessels, as shown in figure \ref{position}b, which increases the cell cycle entry. In clinical practice, among the factors that have been reported as being associated with early or late recurrence, there are characteristics such as tumor size and number of tumors before PH, micro and macrovascular invasion, degree of tumor differentiation, alpha-fetoprotein levels \cite{c21,c22,c23,c24,c25,c26,c27} but how tumor location in the liver will influence the HCC recurrence has been poorly described and with heterogeneous results \cite{c28,c29,c210,c211}.
	
\begin{figure}[!htb]
		\centering
		\includegraphics[width=0.9\linewidth]{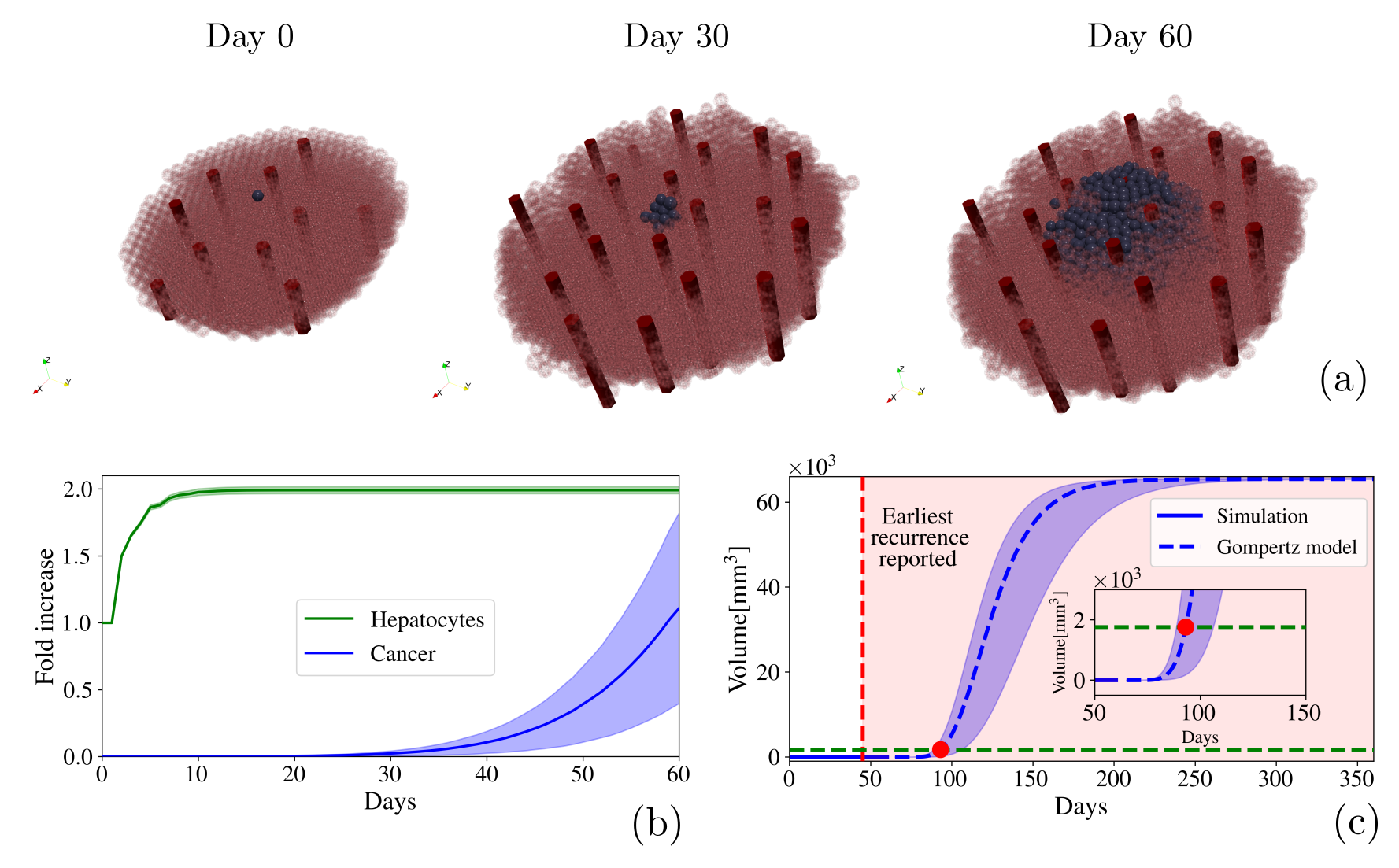}
		\caption{\textit{Hepatocellular carcinoma} \textbf{(a)} Schematic representation of the recurrence of a hepatocellular carcinoma after a $70\%$ PH. Hepatocytes are represented by transparent pink cells to have a better view of the blood vessels (red tubes) and the tumor (blue cells) growth \textbf{(b)} Fold-increase in liver regeneration and hepatocellular carcinoma recurrence. \textbf{(c)} Time based volume estimation of the hepatocellular carcinoma by using simulation data (blue line) to adjust a Gompertz model growth curve (blue dotted line). Red dotted line represents the earliest recurrence reported in clinical observations and the red point along with the green dotted line,represents the minimum volume that the tumor must reach to be detected ($5$ mm) by imaging diagnose.}
		\label{cancer}
	\end{figure}

	We have considered the mean value of our simulations and analyze the recurrence dynamics. We observe that, despite the initial location, there is a delay in the cancer cell reactivation in comparison to the liver regeneration process. Figure \ref{cancer}a shows the process in a qualitative fashion. Cancer cells start growing after the liver finishes its regeneration process. In this case, they grow inwards, following the increase of oxygen concentration, and consequently they proliferate towards the blood vessels (the animation of the HCC recurrence can be seen at the S3 Video.). The growth of the tumor cells showed in figure \ref{cancer}b, allowed us to compute the specific growth rate of the tumor (eq. \ref{sgr}), which give us $\alpha=0.053 \;\%/\text{day}$. By using the Gompertz Model (eq. \ref{gompertz}), we can predict the tumor growth kinetics. We have considered the carrying constant $K$ as a $50$ mm tumor, based on the biggest tumor size in the \textit{Milan criteria} \cite{milan}. The initial volume would be the tumor size at the end of the simulation which was $V_{0}=0.0028\: \text{mm}^{3}$, and the constant $\alpha$ is the specific growth rate. The prediction of Gompertz model is shown in figure \ref{cancer}c as a blue dotted line. If we consider a detection size of $5$ mm, the recurrence of the modeled tumor could be detected around the $95^{th}$ day (as shown in the inset of figure \ref{cancer}c). In figure \ref{cancer}b, shaded regions represent the standard deviations of $40$ simulations for liver regeneration and for tumor growth, while in figure \ref{cancer}c shaded regions represent the Gompertz model calculated based on the standard deviation values. This is in qualitative agreement with the clinical observations presented in Refs. \cite{recurrencia1,recurrencia2,recurrencia3}, in which the earliest time from surgery to first recurrence was $45$ days. It is considered an early recurrence.
	
\begin{figure}[!htb]
		\centering
		\includegraphics[width=\linewidth]{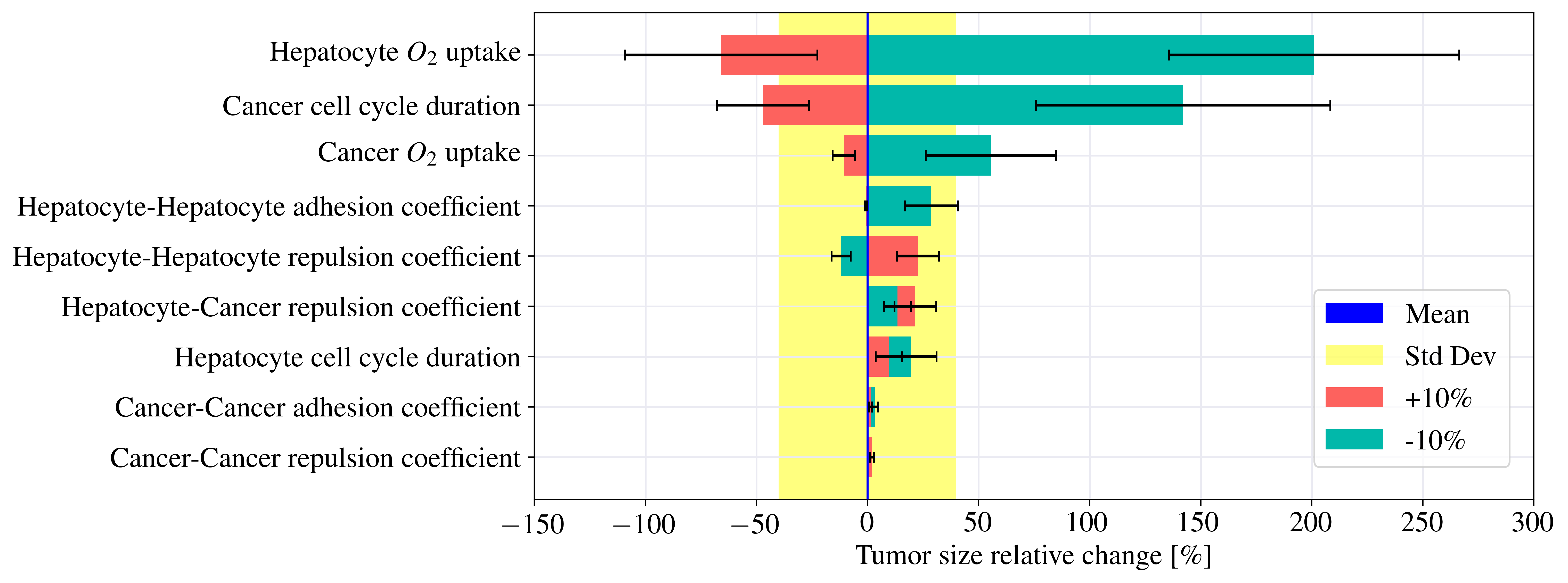}
		\caption{\textit{Sensitivity study.} Relative change of the tumor final size, based on the variation of the parameters that feed our model. The blue line represents the mean tumor size, the yellow shaded region represents the standard deviation of the mean value based on the stochasticity of the model. Red bars and blue bars represent the mean value of the tumor growth when the parameter original value is increased and decreased by $10\%$ respectively and the error bars represent the standard deviation. The three parameters most likely to make an impact on the tumor growth are: the hepatocyte oxygen uptake constant, cancer cells cycle duration and cancer cells oxygen uptake constant.}
		\label{sensitivity}
	\end{figure}

	Our model has thus proven its ability to estimate the growth kinetics of the tumor based on its early stage growth. It could well turn into a useful tool to determine the optimal follow-up interval after the PH. Currently, the American Association for the Study of Liver Diseases, the European Association for the Study of Liver, and the Asian Pacific Association for the Study of the Liver recommend that HCC screening must be conducted at $6$-month intervals \cite{sgr22}. But a consensus interval for recurrence after surgical resection has not been established. Moreover, this recommendation is based on the assumption that the HCC growth rate is similar in every patient. However, tumor growth in general is strongly affected by the microenvironment \cite{sgr213}, and HCC growth rate is likely affected by host factors such as age, sex, preexisting diseases, etc., as well as by tumor factors, such as initial average HCC diameter, tumor multiplicity, etc. \cite{hccsgr2}. Therefore, we used our model to establish a proof of concept of how knowing different parameters that determine tumor development can predict such behavior. We have performed an exploratory sensitivity analysis by varying $\pm10\%$ the input variables that feed our model: oxygen uptake of tumor cells and hepatocytes, hepatocytes and cancer cell cycles duration, repulsion and adhesion coefficients between cancer cells and hepatocytes. Figure \ref{sensitivity} shows the tumor size relative change based on the variation of those parameters. The blue line represents the mean tumor size, the yellow shaded region represents the standard deviation of the mean value based on the stochasticity of the model. Red bars and blue bars represent tumor growth when the parameter original value is increased and decreased by $10\%$ respectively. The parameters whose bars fall into the yellow zone do not modify the tumor growth, hence, according to our model only three parameters mostly control tumor growth. The largest influence is exerted by the hepatocyte oxygen uptake constant. When we increase this constant by $10\%$, the hepatocytes need more oxygen to keep metabolically active. That causes a reduction in the oxygen concentration in the microenvironment which leads to a reduction in the cancer cell cycle entry. It shrinks the tumor final size by $66\%$. On the other hand, when the hepatocyte oxygen uptake constant is decreased by $10\%$, there is more oxygen available in the microenvironment, and the final tumor size will be increased by $200\%$. Reducing the same constant in cancer cells, also impacts in the final tumor size, however, not as significantly as in the previous instance. The second most influential parameter is the cancer cell cycle duration. A $10\%$ reduction leads to an increase of the tumor final size by $140\%$. Conversely, increasing this parameter by the same amount will shrink the tumor size a $47\%$.
	
	In clinical practice, those parameters can be associated to patient specific factors. It has been reported that cell division rates decrease with age \cite{divcel, divcel2}, so that younger individuals have a higher fraction of dividing cells than older individuals. In fact, although the removal of up to $75\%$ of the total liver volume is feasible in a young patient ($\leq40$ years of age) with normal hepatic parenchyma, it is suggested that resection must be more conservative in elderly patients (e.g., $\geq70$ years of age) \cite{clavien}. Other studies show that regenerating liver consumed an increased amount of oxygen, especially during DNA synthesis after hepatectomy \cite{50ph, oxigeno}, that is why dynamic assessment of preoperative exercise capacity may be a useful predictor of short- and long-term postoperative prognosis \cite{ejercicio}. Even though at its current stage, our model can qualitatively and quantitatively capture characteristics of the processes of liver regeneration and  hepatocellular carcinoma recurrence, the data from markers that assess the patient's liver overall health could be used to feed the algorithm with specific patient factors, and in this way the model would easily transform into a workbench for hypotheses testing.

	\section{Discussion}
	\label{sec:discussion}
	
	In this study, we developed a 3-D off-lattice agent-based model to simulate large multi-cellular systems with the ability to test tissue-scale effects resulting from cell-scale interactions. One of the main characteristics of this type of models is their high predictive power. Obviously, in this context, a pre-requisite is a careful calibration and validation of the ABM. This means, one should first tune the model with real data until the known tissue behavior is reproduced. Since the liver has a remarkable capacity to regenerate, liver regeneration after partial hepatectomy has long been an excellent testing ground for modeling tissue regeneration. Moreover, although the liver consists of various types of cells, hepatocytes account for about $80\%$ of liver weight and about $70\%$ of all liver cells, which simplifies some of physiological aspects the model has to address. One can focus only on hepatocytes in order to study the relation of organ size with number and size of cells, and this makes them the ideal candidates to be the agents of our model. On the other hand, taking into consideration that the liver has a double supply of oxygen through the portal vein and hepatic artery, intrahepatic blood flow is highly regulated both by the disposition of its anatomical unit, which is the lobule, as well as by the interaction of its components with the extracellular matrix. Dirichlet nodes turn out to be the perfect computational construct to represent these blood vessels' architecture and function, and thus add further realism to the oxygen diffusion within the model organ. 
	
	It is to be stressed that the precise  mechanisms of initiating, promoting and terminating regenerative responses remain to some extent unknown. Here we have proposed a computational regulatory mechanism based on a substrate that diffuses in the cell microenvironment, substrate here denoted as growth factor. We assume that when the liver is intact, the microenvironment is in homeostasis and, although they are metabolically active, hepatocytes remain dormant in the cell cycle. When a partial hepatectomy takes place, each cell secretes a constant amount of growth factor and, depending on its concentration, they will undergo either hypertrophy or proliferation, as observed by Miyaoka et al. With all that information, we have performed two experiments to calibrate and validate our model. First we studied the liver regeneration process after a $30\%$ PH on an normal hepatic parenchyma. In this context, the liver recovers its volume due to the hypertrophy of the hepatocytes, no proliferation is observed. Secondly, we studied the liver regeneration process after a $70\%$ PH. Here hepatocytes enlarge their volume but also proliferate. The outcomes of our simulations, which are shown in section \ref{sec:results;subsec:regeneration}, are in agreement with the observations reported by Miyaoka et al.
	
	We consider important to stress the fact that despite the major simplifications that our ABM presents, its outcomes show that the overall functionality of the model is preserved. Since the liver does not recover its original shape, the reduced spherical model served properly to reproduce the liver regeneration dynamics. Moreover, since the grid we used to simulate the liver is orders of magnitude smaller than an actual liver, the region could be considered as representative of sub-regions of the liver, avoiding architectural issues. On the other hand, our results show that by tuning the effective oxygen diffusion and decay constants, we can reproduce the effect of sinusoids in the liver lobules. Although our model does not place limitations on the liver size and shape, and allow us to introduce some inhomogeneities in the microenvironment in order to simulate the sinusoids, we believe that it will increase the computational burden of the simulations and it will not make much of a difference to the results. The comparison between \textit{in silico} and \textit{in vitro} systems shows that even a model based on simple rules governing cell cycle, intercellular bonding and basic physical relationships between neighbouring cells can successfully reproduce the behaviour of a real biological system.
	
	Once the model is calibrated, it can be used to study the emergent behavior of the tissue in different scenarios. First, we implemented our model to test another degree of PH. Since $50\%$ PH has now been successfully and frequently used for surgical removal of liver associate tumors, we have performed an \textit{in silico} $50\%$ PH. Here, similar to the $70\%$ PH, hepatocytes enlarge their volume and also proliferate but with a reduced replication rate. We adjusted the results of liver regeneration time and percentage of proliferation based on the degree of PH of our simulated PH, to get an approximation of the liver behavior under other resection degrees.
	
	Second, since the PH is the first line of treatment for patients with hepatocellular carcinoma in stage 0-A of BCLC staging system without clinically significant portal hypertension, we used our model to predict the potential recurrence of the tumor in the remnant liver after a $70\%$ PH. The outcome of our simulations is in accordance with clinical observations, which comes to reinforce our confidence in the applicability of this approach in other scenarios. In that context and based on the fact that HCC growth rate is likely affected by host factors, we have performed an exploratory sensitivity analysis by varying $\pm10\%$ the input variables that feed our model. We found that there are three parameters that according to our model most likely make an impact on the tumor growth: the hepatocyte oxygen uptake constant, cancer cells cycle duration and cancer cells oxygen uptake constant. We take this as a proof of concept of how knowing different parameters that determine tumor development predict such behavior in this 3-D off-lattice agent-based model. We have also studied the tumor growth rate based on the initial position of the the residual tumor clone. We observed that tumor volume vary depending on the location of the residual tumor clone. It grows faster when is located at the periphery of the liver. This result suggest that, although the geometry of the liver does not influence when study liver regeneration, it might make an impact on the HCC recurrence.
	
	It is important to mention that even though at its current stage, our model can qualitatively and quantitatively capture characteristics of the processes of liver regeneration and  hepatocellular carcinoma recurrence, it is not calibrated to any particular type of patient specific parameters (age, sex, pre-existing conditions,...). This is obviously a handicap for the model's direct application in clinical practice. However, the algorithm can be fed with specific patient factors, and in this way the model would easily transform into a workbench for hypotheses testing. The data required for model calibration should include tissue biopsies, as well as data from markers that assess the patient's liver overall health. Such data set can be used to both quantify different cell types, and record their spatial arrangement. Half the images can be considered training data and be used to calibrate the ABM, while the other half can serve as testing data reserved to check if the model predicts tissue behavior with reasonable accuracy. In a way, this procedure resembles the training of a  neural network. 
	
	Some limitations should be stressed in  our current model. Firstly, the sample sizes are considerable smaller than those expected in real situations. As mentioned, in order to accommodate the simulations to temporal and spatial scales accessible to regular computer facilities, the lattice we used to model liver regeneration is $1\; \text{mm}^3$ in volume, far too small when compared with actual human liver size. Even if the region could be considered as representative of sub-regions of the liver, is obvious that important size effects might be left out. This challenge can be tackled resorting to new programming methodologies based on massively parallel computational approaches for multi-core and multi-tensor core devices such as the modern general purposes graphic processing units. Research along these lines is currently in progress. 
	
	A second limitation at the cellular-tissue scale is that some important cell types found in the liver, such as, biliary epithelial cells, stellate cells and Kupffer cells, are not included in the current model. Even though hepatocytes are responsible for most of the metabolic and synthetic functions of the liver, a future improvement could be the implementation of the interaction between  different cell types and explore how it affects liver regeneration processes. A substantial amount of physiological research would be required as a prior step to the implementation of this model upgrade. 
	
	Finally, a few words concerning our model's implementation. We have been particularly careful to construct a software  platform  in a modular and extensible fashion. The aforementioned modules can be replaced with ones with more fine-grained versions as discussed, so that more specific details can be incorporated (as properties) and new processes as well (as methods) with different degrees of detail. Even though our model is not a $1:1$ \textit{in silico} copy of the liver and, therefore, it can not accurately describe in full detail the complex biology of liver regeneration and HCC recurrence, it could serve as a tool to test different hypotheses, as well as for testing and analyzing possible outcomes using multiple plausible parameter combinations. We are confident that once the goal of implementing patient specific factors is reached and the model undergoes a rigorous calibration and validation, it could be used as a platform for \textit{in silico} conducting virtual clinical trials.

	\section{Supporting information}
	\label{sec:supporting}
	
	\noindent\textbf{S1 Text. Supplementary information.} Extensive supplementary information including full mathematical model details, supporting literature and reference parameter values for main examples.
	
	\noindent\textbf{S1 Video. Liver regeneration after 30\% PHx.} 3D simulation of the liver regeneration after a 30\% partial hepatectomy. Video available at \url{https://youtu.be/7URQweHJ6Tk}
	
	\noindent\textbf{S2 Video. Liver regeneration after 70\% PHx.} 3D simulation of the liver regeneration after a 70\% partial hepatectomy. Video available at \url{https://youtu.be/xFtj6ZjrsKk}
	
	\noindent\textbf{S3 Video. Hepatocellular carcinoma recurrence after a 70\% PHx.} 3D simulation of the hepatocellular carcinoma recurrence after a 70\% partial hepatectomy. Video available at \url{https://youtu.be/znVfIxt_Gno}.
	
	\noindent\textbf{Source code.} The code used for running experiments is available at \url{https://github.com/lmluque/abm}

	\section{Acknowledgements}
	L.M.L. would like to thank MD Julieta Cipollone for useful discussions.
	
	\section{Author Contributions}
	\label{sec:contributions}
	
	\noindent\textbf{Conceptualization:} Luque, Llamoza Torres, Carlevaro.
	
	\noindent\textbf{Data curation:} Luque.
	
	\noindent\textbf{Formal analysis:} Luque, Lomba, Carlevaro.
	
	\noindent\textbf{Funding acquisition:} Lomba.
	
	\noindent\textbf{Investigation:} Luque.
	
	\noindent\textbf{Methodology:} Luque, Carlevaro.
	
	\noindent\textbf{Project administration:} Lomba.
	
	\noindent\textbf{Resources:} Lomba, Carlevaro.
	
	\noindent\textbf{Software:} Luque.
	
	\noindent\textbf{Supervision:} Lomba.
	
	\noindent\textbf{Validation:} Luque.
	
	\noindent\textbf{Visualization:} Luque.
	
	\noindent\textbf{Writing-Original draft preparation:} Luque.
	
	\noindent\textbf{Writing-Review \& editing:} Llamoza Torres, Lomba, Carlevaro.

	\section{References}
	\label{sec:references}

\end{document}